\documentclass[12pt,english,titlepage]{article}
\usepackage{ae}
\usepackage[T1]{fontenc}
\usepackage[latin9]{inputenc}
\usepackage{geometry}
\usepackage{textcomp}
\usepackage{setspace}
\usepackage{amsmath}
\usepackage{amssymb}
\usepackage{pgf}
\usepackage{tabularx}
\usepackage[super]{cite}

\makeatletter

\newcommand{\dpr}{^{\prime\prime}}
\def\vec#1{\ensuremath{\mathchoice{\mbox{\boldmath$\displaystyle#1$}}
{\mbox{\boldmath$\textstyle#1$}}
{\mbox{\boldmath$\scriptstyle#1$}}
{\mbox{\boldmath$\scriptscriptstyle#1$}}}}

\def\tens#1{\vec{#1}}
\def\max{\textrm{max}}
\def\min{\textrm{min}}

\usepackage{babel}
\makeatother

\begin{document}
\begin{center}
{\Large Three-wave interactions of dispersive plasma waves propagating parallel
to the magnetic field}
\par\end{center}{\Large \par}

~

~

\begin{center}
F. Spanier
\par\end{center}

\begin{center}
Institut für Theoretische Physik und Astrophysik, Universität Würzburg, Germany
\par\end{center}

\begin{center}
+49(931)888-4932; fax: +49(931)888-4932; email: fspanier@astro.uni-wurzburg.de
\par\end{center}

\begin{center}
and
\par\end{center}

\begin{center}
R. Vainio$^*$
\par\end{center}

\begin{center}
Department of Physics, University of Helsinki, Finland
\par\end{center}

\begin{center}
+358(9)191-50676; fax: +358(9)191-50610; email:
rami.vainio@helsinki.fi
\par\end{center}

~

\begin{center}
submitted to Adv. Sci. Lett.\thispagestyle{empty}
\par\end{center}

\begin{abstract}
\begin{onehalfspace}
Three-wave interactions ($\mbox{M}\leftrightarrow\mbox{P}+\mbox{Q}$)
of plasma waves propagating parallel to the mean magnetic field at
frequencies below the electron cyclotron frequency are considered. We consider
Alfvén--ion-cyclotron waves (L), fast-magnetosonic--whistler waves
(R), and ion-sound waves (I). Especially the weakly
turbulent low-beta plasmas like the solar corona are studied, using the cold-plasma dispersion relation for
the transverse waves (L, R) and the fluid-description of the warm
plasma for the longitudinal waves (I). We analyse the resonance
conditions for the wave frequencies $\omega$ and wavenumbers $k$
(i.e., $\omega_{\rm M}=\omega_{\rm P}+\omega_{\rm Q}$ and $k_{\rm
  M}=k_{\rm P}\pm k_{\rm Q}$, and the interaction rates $u^{\rm MPQ}$ of the waves for
all possible combinations of the three wave modes, and list those
reactions that are not forbidden. One of the waves has to be
longitudinal and two transverse.  This demonstrates an
attractive feature of the theory: the conservation of angular momentum is implicitly built in. In a low-beta plasma, non-zero reaction rates are obtained for
(i) $\mbox{L}^+ \leftrightarrow \mbox{I}^++\mbox{L}^-$ and $\mbox{R}^+
\leftrightarrow \mbox{I}^+ + \mbox{R}^-$ in a wide frequency range
extending from the MHD frequency range to the resonances of the waves;
(ii) for $\mbox{L}^+ \leftrightarrow \mbox{L}^++\mbox{I}^+$ in more
narrow frequency range, where at least one of the L waves is in the
dispersive frequency range; and (iii) $\mbox{I}^+ \leftrightarrow
\mbox{L}^+ + \mbox{R}^\pm$ close to the resonance of the L mode. The
reaction types (ii) and (iii) have, to our knowledge, not been
discussed before in low-beta plasmas. In high-beta plasmas, reactions
$\mbox{I}^+ \leftrightarrow \mbox{R}^+ + \mbox{L}^-$ and $\mbox{I}^+
\leftrightarrow \mbox{L}^++\mbox{R}^-$ are the main reactions,
extending down to the MHD frequencies and discussed earlier for the
non-dispersive case, but new reactions involving dispersive waves are
found in high-beta plasmas as well: $\mbox{I}^+ \leftrightarrow
\mbox{L}^++\mbox{R}^+$ and $\mbox{R}^+ \leftrightarrow \mbox{I}^+ +
\mbox{R}^\pm$ are now possible in limited frequency ranges involving at
least one dispersive transverse wave. We discuss the implications of
the discovered new reactions to turbulent cascading in space
plasmas.
\end{onehalfspace}

\end{abstract}

\section{Introduction}

Three-wave interactions constitute the lowest-order non-linear
coupling between wave-modes in plasmas. These interactions are of
form, where either two modes coalesce into one or one mode decays into
two, i.e., $\mbox{M}\leftrightarrow\mbox{P}+\mbox{Q}$, with momentum
and energy being conserved at microscopic level, i.e., $\omega_{{\rm
    M}} = \omega_{{\rm P}}+\omega_{{\rm Q}}$ and $\vec{k}_{{\rm M}} =
\vec{k}_{{\rm P}}+\vec{k}_{{\rm Q}}$. Thus, three-wave interactions
provide the leading-order description of the evolution of weak
turbulence, where it is assumed that the amplitudes of the interacting
wave packets are small enough for a perturbation theory approach to
apply. Turbulence evolution, on the other hand, is one of the key
issues in plasma astrophysics and space physics. It is intimately
related to fundamental questions like plasma heating and particle
acceleration and transport in collisionless plasmas. To get started
with weak turbulence theory, the reader is referred to \cite{M86}.

In this paper we will investigate three-wave interactions of plasma waves
propagating parallel to the mean magnetic field in a magnetised, low-beta
plasma. The pioneering work by Chin and Wentzel \cite{CW72} considered
three-wave interactions between low-frequency plasma waves under the
random-phase approximation, assuming that the waves fulfill the MHD dispersion
relations, i.e., $\omega=v_{{\rm A}}k_{\parallel}$ for the transverse waves and
$\omega=v_{{\rm S}}k_{\parallel}$ for the longitudinal waves. Their study
revealed that for a low-beta plasma, three-wave interactions where an Alfvén
wave decays into a parallel-propagating sound wave and an
anti-parallel-propagating Alfvén wave, i.e.,
$\mbox{A}^+\leftrightarrow\mbox{S}^++\mbox{A}^-$, dominate the evolution of
weak turbulence. In these interactions, the number of Alfvén wave quanta is
conserved, and since the sound waves are generally easily damped, the
interactions proceed typically in the direction of wave decay. Thus, their
study predicted an inverse cascade of energy from high to low frequencies with
a simultaneous heating of the plasma by the dissipation of the sound waves
emitted by the three-wave interaction. Their model has been applied by numerous
authors to study different astrophysical scenarios from the heating of the
solar corona \cite{W74} to the confinement and transport of cosmic rays (CR) in
the galaxy \cite{S75} and more recently to CR acceleration in astrophysical
shocks \cite{VS05}.  The work of Chin and Wentzel, however, was developed under
the MHD approximation. In this work, we will consider an extension to the work
by adopting dispersion relations more suitable for a collisionless plasma
\cite{M86,LM06}. Specifically, we will consider interactions between left- and
right-hand circularly polarised transverse waves and longitudinal ion sound
waves at frequencies well below the electron gyro-frequency. We will apply the
results of the analysis to conditions in the solar corona, pointing out several
new evolution scenarios of parallel propagating plasma waves, which have
relevance to, e.g., coronal heating and particle acceleration at the Sun. This treatment is considered a kinetic treatment for two major reasons: First of all we are not using the limited MHD dispersion relations which would not hold for frequencies well above the ion gyrofrequency. Secondly we are not bound to thermal distribution functions of the background plasma, which means we do not depend on collisions to produce thermal distributions. As will be shown later in this paper for Maxwellian distributions, however, the actual interactions only depend on the parameter $\beta$ and the ratio $\Omega_\textrm{i}/\omega_\textrm{pi}$.

The structure of the paper is as follows: in \S 2 we will summarise
the theory of three-wave interactions for collisionless plasmas, in \S
3 we will analyse in detail the three-wave interactions between plasma
waves propagating parallel to the magnetic field, in \S 4 we will
discuss the results and present the conclusions from the analysis.

\section{Three-wave interactions in a collisionless plasma}

The framework used here to describe the interaction of waves in
collisionless plasmas employs a quasi-particle description of waves
derived previously by \cite{CW72} and described in detail by
\cite{M86} and references therein. These theories are, as already
mentioned above, usually derived for a fluid plasma (for \cite{CW72})
or an unmagnetised plasma (as in \cite{M86}).

Our calculations in the kinetic limit for magnetised plasmas follow
\cite{MS72a} using the interaction tensor from \cite{MS72b}. The physical picture is the following: The interaction of waves is transmitted by the electric field of these waves. From this interaction an infinite series of interaction terms is given, where the linear (first order) term describes the propagation and the first nonlinear (quadratic) term is the simplest form of interaction.

Given three reservoirs of wave quanta denoted by $N^\mathrm{M}$,
$N^\mathrm{P}$, and $N^\mathrm{Q}$ the evolution equation for the
occupation number of the wave quanta is given by
\begin{eqnarray}
    \frac{\partial N^\mathrm{M}(\vec{k})}{\partial t} &=& \int \frac{\mathrm{d}^3 k^\prime}{(2\pi)^3}\int \frac{\mathrm{d}^3
    k^{\prime\prime}}{(2\pi)^3}
    u^\mathrm{MPQ}(\vec{k},\vec{k}^\prime,\vec{k}^{\prime\prime})\nonumber\\
    && \times \left\{N^\mathrm{P}(\vec{k}^\prime)N^\mathrm{Q}(\vec{k}\dpr)
    - N^\mathrm{M}(\vec{k})[N^\mathrm{P}(\vec{k}^\prime) + N^\mathrm{Q}(\vec{k}\dpr)]\right\}\\
    \frac{\partial N^\mathrm{P}(\vec{k}^\prime)}{\partial t} &=& -\int \frac{\mathrm{d}^3
    k}{(2\pi)^3}\int \frac{\mathrm{d}^3
    k^{\prime\prime}}{(2\pi)^3}
    u^\mathrm{MPQ}(\vec{k},\vec{k}^\prime,\vec{k}^{\prime\prime})\nonumber\\
    && \times \left\{N^\mathrm{P}(\vec{k}^\prime)N^\mathrm{Q}(\vec{k}\dpr)
    - N^\mathrm{M}(\vec{k})[N^\mathrm{P}(\vec{k}^\prime) + N^\mathrm{Q}(\vec{k}\dpr)]\right\}\\
    \frac{\partial N^\mathrm{Q}(\vec{k}^{\prime\prime})}{\partial t} &=& -\int \frac{\mathrm{d}^3
    k}{(2\pi)^3}\int \frac{\mathrm{d}^3
    k^{\prime}}{(2\pi)^3}
    u^\mathrm{MPQ}(\vec{k},\vec{k}^\prime,\vec{k}^{\prime\prime})\nonumber\\
    && \times \left\{N^\mathrm{P}(\vec{k}^\prime)N^\mathrm{Q}(\vec{k}\dpr)
    - N^\mathrm{M}(\vec{k})[N^\mathrm{P}(\vec{k}^\prime) + N^\mathrm{Q}(\vec{k}\dpr)]\right\}.
\end{eqnarray}
Here, the primed [double-primed] quantities refer to the mode P[Q] and
the unprimed quantities to mode M. The interaction rate is defined as
\begin{eqnarray}
    u^\mathrm{MPQ}(\vec{k},\vec{k}^\prime,\vec{k}^{\prime\prime})
    &=& \frac{8(2\pi)^7 \hbar c^4}{\omega^\mathrm{M}
    \omega^\mathrm{P}\omega^\mathrm{Q}}
    \frac{W_E^\mathrm{M}}{W_\mathrm{T}^\mathrm{M}}
    \frac{W_E^\mathrm{P}}{W_\mathrm{T}^\mathrm{P}}
    \frac{W_E^\mathrm{Q}}{W_\mathrm{T}^\mathrm{Q}} \times \nonumber\\
    && \vert
    \kappa^\mathrm{MPQ}(\vec{k},\vec{k}^\prime,\vec{k}^{\prime\prime})\vert^2
    \delta^3(\vec{k}-\vec{k}^\prime-\vec{k}^{\prime\prime})\delta(\omega - \omega^\prime - \omega\dpr)
\end{eqnarray}
The interaction can be derived when using the vector potential of the electric field (in temporal gauge). From the vector potential of all involved waves the response tensor can be calculated, which is basically given by the quantity $u^\mathrm{MPQ}$. In weak turbulence theory the response tensor is calculated only to the second order in the vector potentials. The fact that the electric fields are interacting is resembled by the factor $W_E/W_\mathrm{T}$, which simply describes the ratio of electric to total energy. The physical details of the response tensor are now hidden in $\kappa^{\mathrm{MPQ}}$:
\begin{eqnarray}
    \kappa^{\mathrm{MPQ}}(\vec{k},\vec{k}^\prime,\vec{k}^{\prime\prime})
    &=& (e_i^\mathrm{M})^* e_j^\mathrm{P} e_l^\mathrm{Q}
    \kappa_{ijl}(\vec{k},\omega;\vec{k}^\prime,\omega^\prime;\vec{k}^{\prime\prime},\omega\dpr)\\
    \kappa_{ijl}(\vec{k},\vec{k}^\prime,\vec{k}^{\prime\prime})&=&\frac{1}{2}\left[\tilde\kappa_{ijl}(\vec{k},\vec{k}^\prime,\vec{k}^{\prime\prime})+\tilde\kappa_{ilj}(\vec{k},\vec{k}^{\prime\prime},\vec{k}^\prime)\right]\label{eq:kappa_ijl}
\end{eqnarray}
The quantity $\kappa^\mathrm{MPQ}$ is, thus, a scalar formed by
contracting the three polarisation vectors of the interacting waves
with the quadratic response tensor of the plasma. For our further
calculations we will make use of the quantities $\tilde{\tens\kappa}$
and $\vec{e}$, only. The form of the $\tilde{\tens\kappa}$ tensor is
given by \cite{MS72b}
\begin{eqnarray}
    \tilde\kappa_{ijl} &=& \sum_{\mu,\rho,\nu,\nu^\prime,\nu\dpr}
    \delta_{\mu+\nu^\prime, \rho+\nu+\nu^\prime}\frac{q^3
    n}{c^2}2\pi\int_{-\infty}^{+\infty}\mathrm{d} p_\parallel\int_{0}^{+\infty}p_\perp \, \mathrm{d} p_\perp
    \mathrm{e}^{\mathrm{i} \epsilon (\rho \psi^\prime-\mu\psi)}\times \nonumber\\
    & & \times \mathrm{e}^{\mathrm{i}\epsilon(\nu-\nu^\prime+\nu\dpr)\psi\dpr} J_{\nu\dpr}(z\dpr)
    \frac{V_i(\mu,\vec{k},\vec{v})}{\omega-\mu \Omega-k_\parallel
    v_\parallel}\times \nonumber \\
    & & \times
\left[\alpha_j^*(\omega^\prime,\vec{k}^\prime,\rho)\frac{\partial}{\partial
p_\perp} -\frac{\mathrm{i} \epsilon(\nu-\nu^\prime)}{p_\perp}\beta_j^*
(\omega^\prime,\vec{k}^\prime,\rho)+\gamma_j^*(\omega^\prime,\vec{k}^\prime,
\rho)\frac{\partial}{\partial
    p_\parallel}\right]\times \nonumber\\
    & & \times
\left\{\frac{J_{\nu^\prime}(z\dpr)}{\omega\dpr-\nu\Omega-k_\parallel\dpr
v_\parallel} \left[
\alpha_l^*(\omega\dpr,\vec{k}\dpr,\nu)\frac{\partial}{\partial
p_\perp}+\gamma_l^*(\omega\dpr,\vec{k}\dpr,\nu)\frac{\partial}{\partial
p_\parallel} \right] f(p_\perp,p_\parallel)\right\}\nonumber\\
  \end{eqnarray}

Here, $f$ is the distribution function in momentum space, $n$
is the particle density of the background medium, and $q$ is the
particle charge. We will limit ourselves to the case of gyrotropic
non-relativistic $f$, since relativistic or non-gyrotropic
distribution functions will increase the complexity of the problem
drastically. In the derivation of the equations for the interaction
rates no further assumptions are made. The general form of the
interaction tensor does not depend on the specific form of the
distribution function. It is, however, necessary to choose a specific
form of the distribution function to determine the possible
interactions (which we do in this paper) and to compute the actual
interaction rates from the derived general equations (which we leave
for the future). We use the dispersion relation of ion sound waves
(and partially also those of L- and R-waves) derived from a Maxwellian
distribution function. Using another form of the distribution could,
in principle, change the possible interactions if the resulting
dispersion relations are qualitatively different from those obtained
from the Maxwellian one (e.g., if the assumed monotonic behavior of
$\omega(k)$ would be violated).

The summation of $\mu,\rho,\nu,\nu^\prime,\nu\dpr$  is going from $-\infty$ to $\infty$. $\psi$ is the phase angle of the wave vector around the magnetic field, the
velocity $\vec{V}$ is defined as
  \begin{equation}
    \vec{V}(s,\vec{k},\vec{v}) =\begin{pmatrix}

\frac{v_\perp}{2}[\mathrm{e}^{\mathrm{i}\epsilon\psi}J_{s-1}(z)+\mathrm{e}^{-\mathrm{i}\epsilon\psi}J_{s+1}(z)]\\
-\frac{\mathrm{i}\epsilon
 v_\perp}{2}[\mathrm{e}^{\mathrm{i}\epsilon\psi}J_{s-1}(z)-\mathrm{e}^{-\mathrm{i}\epsilon\psi}J_{s+1}(z)]\\
 v_\parallel
    J_s(z)\end{pmatrix}
  \end{equation}
and the remaining coefficients as
  \begin{eqnarray}
    \alpha_i(\omega,\vec{k},s)&=&\frac{\omega-k_\parallel
    v_\parallel}{v_\perp}V_i(s,\vec{k},\vec{v})-\delta_{i3}(\omega -
s\Omega-k_\parallel
    v_\parallel)\frac{v_\parallel}{v_\perp}J_s(z)\\
    \gamma_i(\omega,\vec{k},s)&=& k_\parallel V_i(s,\vec{k},\vec{v}) +
    \delta_{i3}(\omega -  s\Omega - k_\parallel v_\parallel)J_s(z)\\
    \beta_1(\omega,\vec{k},s)&=&-\frac{\omega-k_\parallel
v_\parallel}{v_\perp}V_2(s,\vec{k},\vec{v})+k_\perp v_\perp \sin\psi\; J_s(z)
\\
    \beta_2(\omega,\vec{k},s)&=&\frac{\omega-k_\parallel
v_\parallel}{v_\perp}V_1(s,\vec{k},\vec{v})-k_\perp v_\perp \cos\psi\; J_s(z)
\\
    \beta_3(\omega,\vec{k},s)&=& \frac{v_\perp k_\perp}{v_\perp}[\sin\psi\;
V_1(s,\vec{k},\vec{v})-\cos\psi\; V_2(s,\vec{k},\vec{v})] \\
    \epsilon &=& \frac{q}{\vert q \vert}\\
    z &=& \frac{k_\perp v_\perp}{\Omega}.
  \end{eqnarray}
When we restrict the discussion to parallel propagating waves, i.e.,
$k_\perp=0$, we obtain $J_s(z)\to \delta_{s0}$ and all terms involving
the phase angles of the waves vanish. Thus, we get
\begin{eqnarray}
\tilde\kappa_{ijl}&=&\sum_{\mu,\rho,\nu=-1}^1\delta_{\mu,\rho+\nu}
 \frac{q^3 n}{c^2}\,2\pi\int_{-\infty}^{+\infty}\mathrm{d}p_\parallel
 \int_0^\infty p_\perp\,\mathrm{d}p_\perp\frac{\tilde{V}_i(\mu,k_\parallel,\vec{v})}
        {\omega-\mu\Omega-k_\parallel v_\parallel}\times\nonumber\\
&&\times\left[
 \tilde\alpha_j^*(\omega^\prime,k_\parallel^\prime,\rho)\frac{\partial}{\partial p_\perp}
 -\frac{\mathrm{i}\epsilon\nu}{p_\perp}\tilde\beta_j^*(\omega^\prime,k_\parallel^\prime,\rho)
 +\tilde\gamma_j^*(\omega^\prime,k_\parallel^\prime,\rho)\frac{\partial}{\partial p_\parallel}
 \right]\times\nonumber\\
&&\times\left\{
 \frac{1}{\omega\dpr-\nu\Omega-k_\parallel\dpr}\left[
  \tilde\alpha_l^*(\omega\dpr,k_\parallel\dpr,\nu)\frac{\partial}{\partial p_\perp}
 +\tilde\gamma_l^*(\omega\dpr,k_\parallel\dpr,\nu)\frac{\partial}{\partial p_\parallel}
  \right] f(p_\parallel,p_\perp)\right\}\nonumber\\
\end{eqnarray}
where
\begin{eqnarray}
\tilde{\vec{V}}(s,k_\parallel,\vec{v})&=&\begin{pmatrix}
    \frac{v_\perp}{2}(\delta_{s,1}+\delta_{s,-1})\\
    -\mathrm{i}\epsilon\frac{v_\perp}{2}(\delta_{s,1}-\delta_{s,-1})\\
    v_\parallel \delta_{s,0}
  \end{pmatrix}\\
\tilde{\vec{\alpha}}^*(\omega,k_\parallel,s)&=&\frac{\omega-k_\parallel v_\parallel}{2}
  \begin{pmatrix}
    \delta_{s,1}+\delta_{s,-1}\\
    \mathrm{i}\epsilon(\delta_{s,1}-\delta_{s,-1})\\
    0
  \end{pmatrix}\\
\tilde{\vec{\beta}}^*(\omega,k_\parallel,s)&=&\frac{\omega-k_\parallel v_\parallel}{2}
  \begin{pmatrix}
    \mathrm{i}\epsilon(\delta_{s,1}-\delta_{s,-1})\\
    \delta_{s,1}+\delta_{s,-1}\\
    0
  \end{pmatrix}\\
\tilde{\vec\gamma}^*(\omega,k_\parallel,s)&=&\begin{pmatrix}
    \frac{k_\parallel v_\perp}{2}(\delta_{s,1}+\delta_{s,-1})\\
    \mathrm{i}\epsilon\frac{k_\parallel v_\perp}{2}(\delta_{s,1}-\delta_{s,-1})\\
    \omega\delta_{s,0}
  \end{pmatrix}.
\end{eqnarray}

We will limit ourselves to propagation of waves parallel and
anti-parallel to the magnetic field. In this case the resonance
conditions for any three-wave interaction M~$\leftrightarrow$~P~$+$~Q
read
\begin{eqnarray}
    k &=& k^\prime \pm k\dpr\label{eq:kres}\\
    \omega &=& \omega^\prime + \omega\dpr\label{eq:wres}
\end{eqnarray}
where the wave numbers, $k=|\vec{k}|$ etc., i.e., the lengths of the
wave vectors $\parallel x_3$, and the frequencies are taken to be
positive. Here, as above, the primed [double-primed] quantities refer
to the mode P[Q] and the unprimed quantities to mode M.

\section{Results}
\subsection{Interaction rates}

The wave modes propagating parallel to the magnetic field can be either purely
transverse (T) or purely longitudinal (Lo). In the first step we will now show,
which of the interactions between these wave modes are forbidden because of
vanishing interaction rates. This can be done by finding the vanishing
components of the $\tilde{\kappa}$ tensor and the symmetries of the
$\tens\kappa$ tensor, which cause the vanishing contraction of the polarisation
tensor $\tens E =\vec{e}^\mathrm{M}\vec{e}^\mathrm{P}\vec{e}^\mathrm{Q}$
(components $E_{ijl}=e^\mathrm{M}_i e^\mathrm{P}_j e^\mathrm{Q}_k$) with the
$\tens\kappa$ tensor.  The results for all possible three-wave reactions of
parallel propagating waves, to be discussed in detail below, are summarised in
Table~\ref{tab:reactions}.

\begin{table}
\caption{Three-wave interactions between waves with wave vectors
  parallel to the ambient magnetic field. Specific dispersion
  relations analysed only at frequencies below the electron cyclotron
  frequency.\label{tab:reactions}}
\begin{tabularx}{\textwidth}{llXl}\hline\hline
     Reaction                                                   & Valid in & Comments & Spin conserved?\\\hline
    $\mathrm{T} \leftrightarrow \mathrm{T} + \mathrm{T}$ & --- & $\tilde\kappa_{ijl}=0$ for all non-zero $E_{ijl}$ & No\\
    $\mathrm{Lo} \leftrightarrow \mathrm{T}+\mathrm{Lo}$ & --- & $\tilde\kappa_{ijl}=0$ for all non-zero $E_{ijl}$ & No\\
    \\
    $\mathrm{L} \leftrightarrow \mathrm{R} + \mathrm{I}$ & --- & terms cancel to produce $E_{ijl}\kappa_{ijl}=0$ & No\\
    $\mathrm{R} \leftrightarrow \mathrm{L} + \mathrm{I}$ & --- & terms cancel to produce $E_{ijl}\kappa_{ijl}=0$ & No\\
    $\mathrm{I} \leftrightarrow \mathrm{L} + \mathrm{L}$ & --- & terms cancel to produce $E_{ijl}\kappa_{ijl}=0$ & No\\
    $\mathrm{I} \leftrightarrow \mathrm{R} + \mathrm{R}$ & --- & terms cancel to produce $E_{ijl}\kappa_{ijl}=0$ & No\\
    \\
    $\mathrm{L}^+ \leftrightarrow \mathrm{L}^+ + \mathrm{I}^-$   & --- & requires $\mathrm{d}\omega_\mathrm{L}/\mathrm{d}k<0$ & Yes\\
    $\mathrm{R}^+ \leftrightarrow \mathrm{R}^+ + \mathrm{I}^-$   & --- & requires $\mathrm{d}\omega_\mathrm{R}/\mathrm{d}k<0$ & Yes\\
    \\
    $\mathrm{R}^+ \leftrightarrow \mathrm{I}^+ + \mathrm{R}^-$   & low $\beta$  & allowed in MHD & Yes\\
                                                                & high $\beta$ & $\omega_\mathrm{R^+}/k_\mathrm{R^+}>c_\mathrm{s}$& \\
    $\mathrm{L}^+ \leftrightarrow \mathrm{I}^+ + \mathrm{L}^-$   & low $\beta$  & allowed in MHD & Yes\\
    \\
    $\mathrm{I}^+ \leftrightarrow \mathrm{L}^+ + \mathrm{R}^-$   & high $\beta$ & allowed in MHD & Yes\\
                                                                & low $\beta$ & $\omega_\mathrm{L}\approx\Omega_\mathrm{i}$\\
    $\mathrm{I}^+ \leftrightarrow \mathrm{R}^+ + \mathrm{L}^-$   & high $\beta$ & allowed in MHD & Yes\\
    \\
    $\mathrm{I}^+ \leftrightarrow \mathrm{L}^+ + \mathrm{R}^+$   & low $\beta$ & $\omega_\mathrm{L}\approx\Omega_\mathrm{i}$ & Yes\\
                                                                & high $\beta$ & $\omega_\mathrm{L}\lesssim\Omega_\mathrm{i}$ & Yes\\
    $\mathrm{L}^+ \leftrightarrow \mathrm{L}^+ + \mathrm{I}^+$   & low $\beta$  & decaying $\omega_\mathrm{L}\lesssim\Omega_\mathrm{i}$ & Yes\\
    $\mathrm{R}^+ \leftrightarrow \mathrm{R}^+ + \mathrm{I}^+$   & high $\beta$ & decaying $\omega_\mathrm{R}\sim\Omega_\mathrm{i}$ & Yes\\\hline
  \end{tabularx}
\end{table}

The components of the $\tilde{\tens\kappa}$ tensor itself vanish whenever
none or two of the indices are equal to 3
\begin{eqnarray}
  \tilde{\kappa}_{111} = \tilde{\kappa}_{112} = \tilde{\kappa}_{121} =
  \tilde{\kappa}_{122} &=& 0 \\
  \tilde{\kappa}_{211} = \tilde{\kappa}_{212} = \tilde{\kappa}_{221} =
  \tilde{\kappa}_{222} &=& 0 \\
  \tilde{\kappa}_{133} = \tilde{\kappa}_{313} = \tilde{\kappa}_{331} &=& 0 \\
  \tilde{\kappa}_{233} = \tilde{\kappa}_{323} = \tilde{\kappa}_{332} &=& 0
\end{eqnarray}
When we take into account that the polarisation vectors of transverse waves
only have non-zero 1 and 2 components, while the longitudinal waves have a
non-zero 3 component, we can make the first statement concerning the wave modes
that may interact. Assuming interactions between three transverse wave modes,
the polarisation tensor's components are zero, whenever one index is 3.
Multiplying this with the $\tens\kappa$ tensor results in a zero. So
interactions between three transverse wave modes will not take place. The same
applies for interactions involving two longitudinal and one transverse wave.

To make further assertions about possible interactions we shall now
investigate the symmetry of the $\tens\kappa$ tensor. The index
symmetries,
\begin{eqnarray}
  \kappa_{113} &=& \kappa_{223}\\
  \kappa_{123} &=& -\kappa_{213}\\
  \kappa_{131} &=& \kappa_{232}\\
  \kappa_{132} &=& -\kappa_{231}\\
  \kappa_{311} &=& \kappa_{322}\\
  \kappa_{312} &=& -\kappa_{321},
\end{eqnarray}
can be calculated either directly or simply by noting that the tensor
is necessarily transversely isotropic, i.e., that there has to be a
rotational symmetry around the $x_3$-axis aligned with the magnetic
field and all the wavevectors, as this is the only preferred direction
in the system.

Before we can determine the components of the polarisation tensor, we have to
specify which wave modes are used. We will now limit our discussion to the L-
and R-mode for the transverse wave fulfilling the dispersion relations
  \begin{equation}
    N^2 = L = 1-\frac{\omega_\mathrm{pi}^2}{\omega(\omega-\Omega_\mathrm{i})}-\frac{\omega_\mathrm{pe}^2}{\omega(\omega+\Omega_\mathrm{e})}
  \end{equation}
and
  \begin{equation}
    N^2 = R = 1-\frac{\omega_\mathrm{pi}^2}{\omega(\omega+\Omega_\mathrm{i})}-\frac{\omega_\mathrm{pe}^2}{\omega(\omega-\Omega_\mathrm{e})}
  \end{equation}
respectively. Here, $N=ck/\omega$ is the refractive index of the wave and $L$
and $R$ are the Stix-parameters \cite{S62}. $\Omega_\mathrm{i}$ and
$\Omega_\mathrm{e}$ are the gyrofrequencies of ions and electrons,
respectively. The plasma frequencies are denoted by $\omega_\mathrm{pi}$ and
$\omega_\mathrm{pe}$. The polarisation vectors associated with these wave modes
are
  \begin{equation}
    \vec{e}^\mathrm{L} = \begin{pmatrix}
      -1/\sqrt{2}\\
      \mathrm{i}/\sqrt{2}\\
      0
    \end{pmatrix}\ , \
    \vec{e}^\mathrm{R} = \begin{pmatrix}
      1/\sqrt{2}\\
      \mathrm{i}/\sqrt{2}\\
      0
    \end{pmatrix}
  \end{equation}
For the longitudinal mode we choose the ion-sound mode
  \begin{equation}
    \omega^2 = k^2 \left(c_\mathrm{s}^2-\omega^2 \lambda_\mathrm{D}^2 \right),
  \end{equation}
where $c_\mathrm{s}=\sqrt{k_\mathrm{B} T_\mathrm{e}/m_\mathrm{i}}$ is the ion
sound speed, $k_\mathrm{B}$ is Boltzmann's constant, $T_\mathrm{e}$ the
electron temperature, $m_\mathrm{i}$ the ion mass and $\lambda_\mathrm{D}$ the
Debye length, with the polarisation vector
  \begin{equation}
    \vec{e}^\mathrm{I} = \begin{pmatrix}
      0\\0\\1
    \end{pmatrix}
  \end{equation}

The non-zero components of $E_{ijl}$ for the interaction pairs of type
$\mbox{T} \leftrightarrow \mbox{T} + \mbox{Lo}$ are given in
Table~\ref{tab:poltens1}.  Combining this with the $\tens\kappa$ tensor, we
find that the interaction rate vanishes identically for the
interactions $\mathrm{L} \leftrightarrow \mathrm{R} + \mathrm{I}$ and
$\mathrm{R} \leftrightarrow \mathrm{L} + \mathrm{I}$, while for the
other two pairs it is
\begin{equation}
  u^{\mathrm{MPQ}} \propto |\kappa_{113} - \mathrm{i} \kappa_{123}|^2
\end{equation}

which is non zero. So for this type of interactions only $\mathrm{L}
\leftrightarrow \mathrm{L} + \mathrm{I}$ and $\mathrm{R}
\leftrightarrow \mathrm{R} + \mathrm{I}$ may have non-zero interaction
rates.

\begin{table}
\begin{center}
  \caption{Non-zero elements of the polarisation tensor
    $E_{ijl}=e_i^\mathrm{M}e_j^\mathrm{P}e_l^\mathrm{Q}$ for
    interactions of the type $\mbox{T} \leftrightarrow \mbox{T} +
    \mbox{Lo}$.}\label{tab:poltens1}
  \begin{tabular}{lrrrr}\hline
    & $E_{113}$ & $E_{123}$ & $E_{213}$ & $E_{223}$ \\\hline\hline
   $\mathrm{L} \leftrightarrow \mathrm{L} + \mathrm{I}$ & $1/2$ & $-\mathrm{i}/2$  & $\mathrm{i}/2$  & $1/2$ \\
   $\mathrm{L} \leftrightarrow \mathrm{R} + \mathrm{I}$ & $-1/2$ &$-\mathrm{i}/2$ &$-\mathrm{i}/2$ & $1/2$ \\
   $\mathrm{R} \leftrightarrow \mathrm{R} + \mathrm{I}$ & $1/2$ &$\mathrm{i}/2$ &$-\mathrm{i}/2$ & $1/2$ \\
   $\mathrm{R} \leftrightarrow \mathrm{L} + \mathrm{I}$ & $-1/2$ &$\mathrm{i}/2$ &$\mathrm{i}/2$ & $1/2$\\\hline
  \end{tabular}
\end{center}
\end{table}

For the next set of interactions, i.e., of type $\mbox{I} \leftrightarrow
\mbox{T} + \mbox{T}$, we have given the non-zero components of the polarisation
tensor in Table~\ref{tab:poltens2}.  As the symmetry of the
$\tens\kappa$-tensor does not change, we find that here the set of allowed
interactions is $\mathrm{I} \leftrightarrow \mathrm{R} + \mathrm{L}$ and
$\mathrm{I} \leftrightarrow \mathrm{L} + \mathrm{R}$.

\begin{table}
\begin{center}
  \caption{Non-zero polarisation tensor elements for interactions of
    the type $\mbox{I} \leftrightarrow \mbox{T} +
    \mbox{T}$.}\label{tab:poltens2}
  \begin{tabular}{lrrrr}\hline
    & $E_{311}$ & $E_{312}$ & $E_{321}$ & $E_{322}$ \\\hline\hline
   $\mathrm{I} \leftrightarrow \mathrm{L} + \mathrm{L}$ & $-1/2$ & $\mathrm{i}/2$  & $\mathrm{i}/2$  & $1/2$ \\
   $\mbox{I} \leftrightarrow \mbox{R} + \mbox{R}$ & $-1/2$ &$-\mathrm{i}/2$ &$-\mathrm{i}/2$ & $1/2$ \\
   $\mbox{I} \leftrightarrow \mbox{R} + \mbox{L}$ & $-1/2$ &$-\mathrm{i}/2$ &$\mathrm{i}/2$ & $-1/2$ \\
   $\mbox{I} \leftrightarrow \mbox{L} + \mbox{R}$ & $-1/2$ &$-\mathrm{i}/2$ &$\mathrm{i}/2$ & $-1/2$\\\hline
  \end{tabular}
\end{center}
\end{table}

Reviewing the set of eliminated interactions one finds that these
selection rules, which were derived solely from the interaction rate
without explicit regard to the resonance conditions, are going back to
angular momentum conservation. When one associates a spin of $+\hbar$
with an L-wave and $-\hbar$ with an R-wave,\footnote{Note, that the
  propagation direction of the wave does not play a role, since the
  plasma physics definition of the handedness of circular polarisation
  is with respect to the magnetic field.} it is obvious, that the
discarded reactions are the ones with non-conserved angular
momentum. All of the remaining interactions are, thus, conserving
angular momentum at the microscopic level. With this reduction of
possible interactions we will now examine, which interactions fulfill
the resonance conditions.

\subsection{Resonance conditions}
Because of the resonance conditions, Eqs.~(\ref{eq:kres}) and (\ref{eq:wres}),
the interactions
\begin{eqnarray}
   \mathrm{L}^+ &\leftrightarrow& \mathrm{L}^+ + \mathrm{I}^-\\
   \mathrm{R}^+ &\leftrightarrow& \mathrm{R}^+ + \mathrm{I}^-,
\end{eqnarray}
can be discarded immediately. The transverse wave has the same
dispersion relation on both sides, and it has
$\mathrm{d}\omega/\mathrm{d}k>0$. However, the frequency of the
transverse wave should be decreasing from left to right in the
reaction while its wavenumber should be increasing, which is not
possible. Thus, although these reactions do conserve the angular
momentum, they cannot occur in plasmas.

The resonance conditions are next carefully evaluated for the
remaining reactions. It is practical to perform this separately in two
distinct parameter regimes: the high and low plasma $\beta$
regions.\footnote{In this paper we define plasma $\beta$ to be the
  ratio of the squared ion-sound speed to the squared Alfv\'en speed,
  $\beta=c_\mathrm{s}^2/V_\mathrm{A}^2$, which is about one half of the
  ratio of gas pressure to magnetic field pressure, if
  $T_\mathrm{e}\gg T_\mathrm{i}$.} This distinction is useful as the
phase speed of the sound wave is below that of the low-frequency
transverse waves in the low-beta case and above it in the high-beta
case at low frequencies.

\subsubsection{Low $\beta$-plasma}
\begin{figure}
\includegraphics[width=0.8\textwidth]{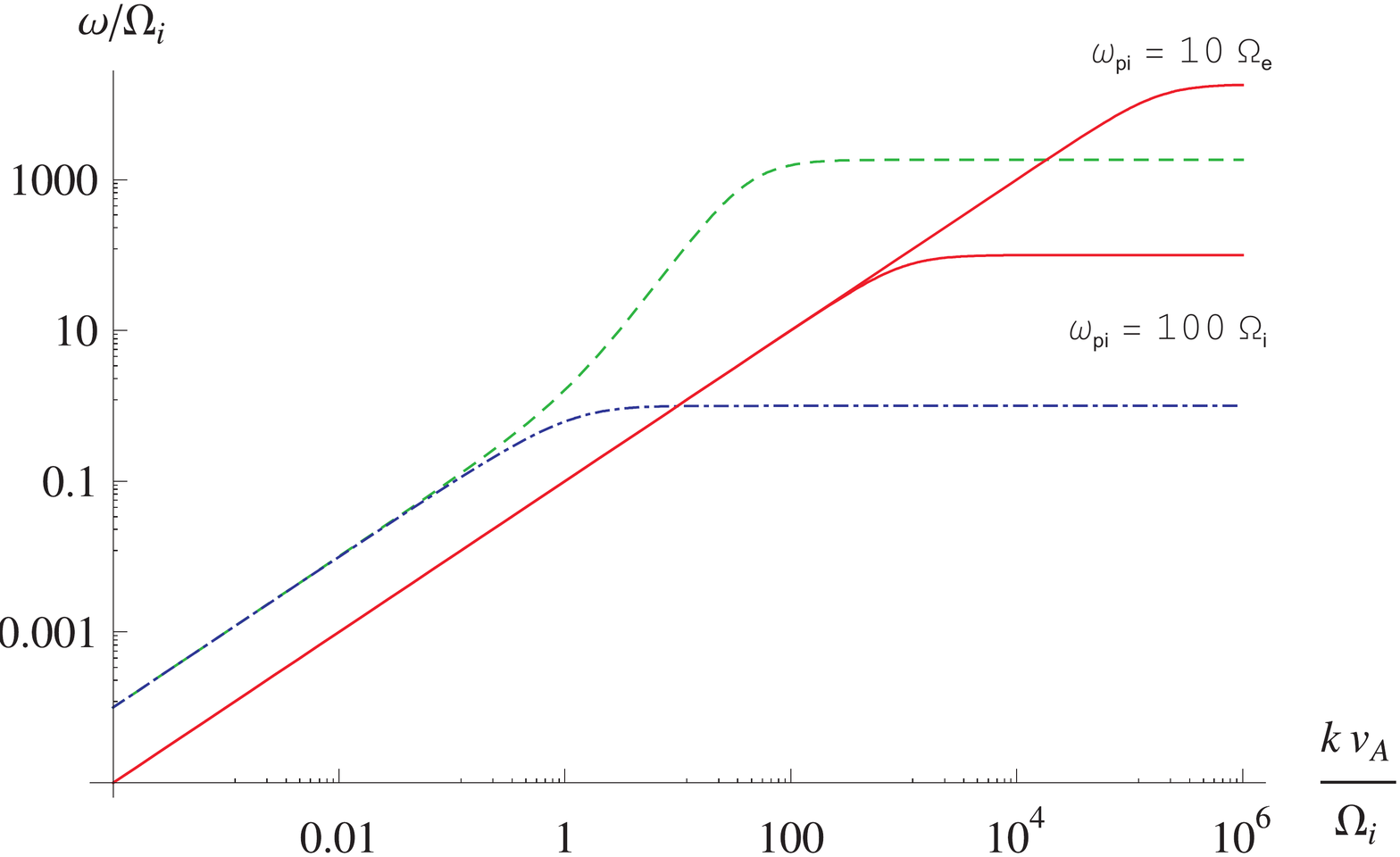}
\caption{Dispersion relation of parallel-propagating plasma waves in
  low-beta plasma, i.e., for $c_\mathrm{s}/v_\mathrm{A}=0.1$. The
  red (solid) curves represent the ion sound wave for a high-density
  ($\omega_\mathrm{pi}=10\,\Omega_\mathrm{e}$) and low-density
  ($\omega_\mathrm{pi}=100\,\Omega_\mathrm{i}$) plasmas, the blue (dot-dashed)
  curve is the L wave and the green (dashed) curve the R wave. Note that
  the horizontal parts of the dispersion relation represent very
  strongly damped waves.}
\label{fig:disprel}
\end{figure}

The dispersion relations of parallel propagating transverse and
longitudinal waves in a low-beta plasma are depicted in
Fig.~\ref{fig:disprel}. In the low-$\beta$ case the frequency of
R-waves well below the electron cyclotron resonance is always higher
than that of the longitudinal wave with the same $k$. This applies
also for the L-wave unless it approaches the ion gyroresonance.

The first two interactions to be inspected are
\begin{eqnarray}
  \mathrm{L}^+ &\leftrightarrow& \mathrm{I}^+ + \mathrm{L}^-\\
  \mathrm{R}^+ &\leftrightarrow& \mathrm{I}^+ + \mathrm{R}^-
\end{eqnarray}
which have been discussed under the MHD approximation already in \cite{VS05}.
They fulfill the resonance condition also in the dispersive limit up to the
wave number, where $\omega_\mathrm{T}(k) \sim \omega_\mathrm{Lo}(k)$, so they
are allowed in the low $\beta$-regime. We will analyse these reactions
numerically below.

The interactions
\begin{eqnarray}
  \mathrm{L}^+ &\leftrightarrow& \mathrm{L}^+ + \mathrm{I}^+\label{LLI}\\
  \mathrm{R}^+ &\leftrightarrow& \mathrm{R}^+ + \mathrm{I}^+\label{RRI}
\end{eqnarray}
can not fulfill the resonance conditions in the MHD regime, since in the
low-frequency range the phase speed of the transverse waves is larger than that
of the longitudinal wave for low $\beta$. The interaction involving the R-wave
may work only if the decaying wave frequency is very close to
$\omega_\mathrm{R}=\Omega_\mathrm{e}$, while the L-wave interaction is limited
to work in a small frequency range close to $\Omega_\mathrm{i}$. We will
analyse the reactions numerically below.

\begin{figure}
  \pgfimage[width=\textwidth]{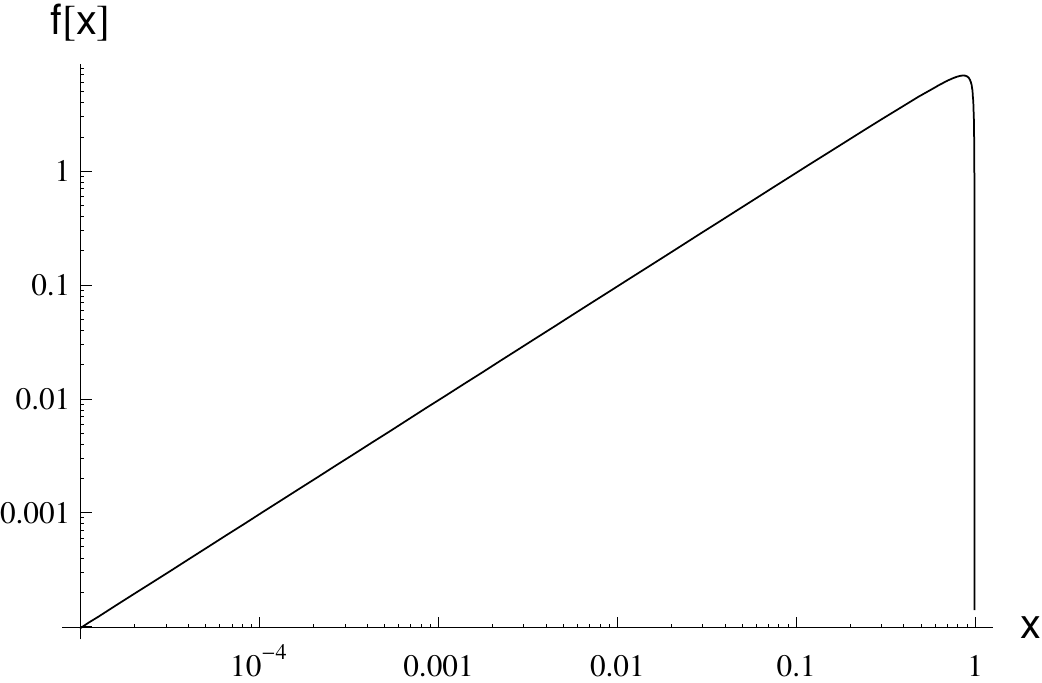}
  \caption{The function of $f(x)$ used to analyse the resonance
    conditions in the interaction
    $\mathrm{L}^+\leftrightarrow\mathrm{L}^++\mathrm{I}^+$. The value
    of $\beta$ used in this plot is $\sim 0.0087$.}
  \label{fig:f}
\end{figure}

The L-wave reaction can be analysed analytically using the approximate
dispersion relations,
\begin{eqnarray}
k_\mathrm{L}(\omega)&\approx& \frac{\omega}{V_\mathrm{A}}
  \sqrt{\frac{\Omega_\mathrm{i}}{\Omega_\mathrm{i}-\omega}}\\
k_\mathrm{I}(\omega)&\approx&\frac{\omega}{c_\mathrm{s}}.
\end{eqnarray}
The resonance conditions yield
\begin{equation}
\frac{\omega}{V_\mathrm{A}}\sqrt{\frac{\Omega_\mathrm{i}}
  {\Omega_\mathrm{i}-\omega}} = \frac{\omega_1}{V_\mathrm{A}}
\sqrt{\frac{\Omega_\mathrm{i}}{\Omega_\mathrm{i}-\omega_1}}+
\frac{\omega-\omega_1}{c_\mathrm{s}},
\end{equation}
where $\omega$ is the frequency of the decaying L wave and $\omega_1$
the frequency of the daughter L wave, $\omega-\omega_1$ being the
frequency of the I wave. This condition can be written in the form
$f(\omega/\Omega_\mathrm{i})=f(\omega_1/\Omega_\mathrm{i})$, where
(Fig.~\ref{fig:f})
\begin{equation}
f(x)=\frac{x}{\sqrt{\beta}}-\frac{x}{\sqrt{1-x}},\quad\beta=c^2_\mathrm{s}/V^2_\mathrm{A}
\end{equation}
The function has two zeros at $x=0$ and $x = 1 - \beta$, and a
positive value in between
and a single maximum.
The values of
$\omega_1$ and $\omega$, thus, correspond to the two roots of the
equation $f(\omega/\Omega_\mathrm{i})=C$, where $C$ is a constant
between 0 and the maximum of $f$
which for a very low beta is at $x\approx
1-(\frac 14 \beta)^{1/3}$. For higher values of $\beta$,
the zero of the derivative, $x_\max$, can be found by iteration taking
$x_\max^{(0)}=1-(\frac 14 \beta)^{1/3}$ and
\begin{equation}
x_\max^{(n+1)} = 1-[\beta(\textstyle{\frac{1}{2}}x_\max^{(n)}-1)^2]^{1/3},
\end{equation}
Thus, one of the
roots (the daughter wave $\omega_1$) is at $0<x<x_\max$ and the other one (the
decaying wave $\omega$) at $x_\max<x<1-c_\mathrm{s}^2/V_\mathrm{A}^2$. The
difference of the roots (multiplied by $\Omega_\mathrm{i}$) gives
$\omega_\mathrm{I}$. This problem may be solved analytically, but the formula is of little practical use.

The R-wave reaction can be analysed in a similar manner, but the details are
left to the interested reader.

By similar arguments, the three interactions
\begin{eqnarray}
  \mathrm{I}^+ &\leftrightarrow & \mathrm{L}^+ + \mathrm{R}^+\label{ILR1}\\
  \mathrm{I}^+ &\leftrightarrow & \mathrm{L}^+ + \mathrm{R}^-\label{ILR2}\\
  \mathrm{I}^+ &\leftrightarrow & \mathrm{R}^+ + \mathrm{L}^-\label{ILR3}
\end{eqnarray}
have no solution for the resonance condition in the low-frequency
limit as the phase speed of the transverse waves exceeds the phase
speed of the longitudinal wave. As above, however, this is no longer
true in the direct vicinity of the resonances of the transverse
waves.

For simplicity, we will consider a high-density plasma, for which a
linear dispersion relation for the longitudinal wave is valid at all
frequencies below $\Omega_\mathrm{e}$.  The resonance conditions for
the first two reactions yield
\begin{equation}
\frac{\omega_\mathrm{L}+\omega_\mathrm{R}}{c_\mathrm{s}}=
 k_\mathrm{L}(\omega_\mathrm{L}) \pm k_\mathrm{R}(\omega_\mathrm{R}),
\end{equation}
i.e.,
\begin{equation}
k_\mathrm{L}(\omega_\mathrm{L}) - \frac{\omega_\mathrm{L}}{c_\mathrm{s}}  =
 \frac{\omega_\mathrm{R}}{c_\mathrm{s}} \mp k_\mathrm{R}(\omega_\mathrm{R}),\label{IC}
\end{equation}
The left-hand side is positive only for $k_\mathrm{L}>
\Omega_\mathrm{i}/c_\mathrm{s}$. The right-hand side is always
positive for the lower sign, interaction (\ref{ILR2}), and for the
upper sign, interaction (\ref{ILR1}), as long as $k_\mathrm{R} <
\Omega_\mathrm{e}/c_\mathrm{s}$. Similarly, for the third interaction
(\ref{ILR3}) one can conclude that solutions are possible only for
$k_\mathrm{R}>\Omega_\mathrm{e}/c_\mathrm{s}$. Thus, the following
reactions are possible if $\omega_\mathrm{pi}>\Omega_\mathrm{e}$:
\begin{eqnarray}
  \mathrm{I}^+ &\leftrightarrow & \mathrm{L}^+ + \mathrm{R}^+\quad\mbox{for }k_\mathrm{L}>\Omega_\mathrm{i}/c_\mathrm{s}\mbox{ and }k_\mathrm{R}<\Omega_\mathrm{e}/c_\mathrm{s}\label{ILR4}\\
  \mathrm{I}^+ &\leftrightarrow & \mathrm{L}^+ + \mathrm{R}^+\quad\mbox{for }k_\mathrm{L}<\Omega_\mathrm{i}/c_\mathrm{s}\mbox{ and }k_\mathrm{R}>\Omega_\mathrm{e}/c_\mathrm{s}\label{ILR5}\\
  \mathrm{I}^+ &\leftrightarrow & \mathrm{L}^+ + \mathrm{R}^-\quad\mbox{for }k_\mathrm{L}>\Omega_\mathrm{i}/c_\mathrm{s}\label{ILR6}\\
  \mathrm{I}^+ &\leftrightarrow & \mathrm{R}^+ + \mathrm{L}^-\quad\mbox{for }k_\mathrm{R}>\Omega_\mathrm{e}/c_\mathrm{s}.\label{ILR7}
\end{eqnarray}
Note that if $\omega_\mathrm{pi}<\Omega_\mathrm{e}$, interactions
involving electron--cyclotron waves, Eqs.~(\ref{ILR5}) and
(\ref{ILR7}), are not possible at all, and the ranges of validity of
the other reactions become modified (i.e.,
$\omega_\mathrm{R}<\omega_\mathrm{pi}-\Omega_\mathrm{i}$ needs to be
satisfied).

In the reactions involving the ion-cyclotron waves, Eqs.\ (\ref{ILR4})
and (\ref{ILR6}), we may write $\omega_\mathrm{L}=\Omega_\mathrm{i}$
and get
\begin{equation}
k_\mathrm{L} =
  k_\mathrm{I}(\omega_\mathrm{R}+\Omega_\mathrm{i}) \mp k_\mathrm{R}(\omega_\mathrm{R}),
\end{equation}
which gives an analytical solution of the wave number of the
left-handed wave as a function of the frequency of the right-handed
wave. Note that this equation is valid also for
$\Omega_\mathrm{i}<\omega_\mathrm{pi}<\Omega_\mathrm{e}$.

In summary, from the seven distinct possible interactions not
eliminated by angular momentum conservation and $\mathrm
{d}\omega/\mathrm{d}k>0$, only two can take place over the full
frequency range in a low $\beta$-plasma with parallel wave propagation:
\begin{eqnarray*}
    \mathrm{L}^+ &\leftrightarrow& \mathrm{I}^+ + \mathrm{L}^-\\
    \mathrm{R}^+ &\leftrightarrow& \mathrm{I}^+ + \mathrm{R}^-.
\end{eqnarray*}
In addition, the interactions (\ref{LLI})--(\ref{RRI}) and
(\ref{ILR1})--(\ref{ILR3}) may have solutions where one of the waves
are in the electron- or ion-cyclotron range. Reactions (\ref{LLI}),
(\ref{ILR1}) and (\ref{ILR2}) all yield solutions where an
ion-cyclotron wave interacts with an MHD wave.

\paragraph{Numerical analysis.}

For the two valid interactions and the one interaction with limited range we
will now show the frequency triads for the resonance condition. As a model
system for a low $\beta$ plasma we have chosen typical parameters for a coronal
hole
\begin{eqnarray*}
  \textrm{Number density} && n_\mathrm{i} = n_e = 2\times 10^7\, \mathrm{cm}^{-3}\\
  \textrm{Electron temperature} && T_\mathrm{e} = 1 \times 10^6\, \mathrm{K}\\
  \textrm{Magnetic field} && B = 2 \,\mathrm{G}
\end{eqnarray*}
The values correspond to $\beta=c_\mathrm{s}^2/V_\mathrm{A}^2\approx
0.0087$ and $\omega_\mathrm{pi}/\Omega_\mathrm{i}\approx 310$.

\begin{figure}
  \pgfimage[width=\textwidth]{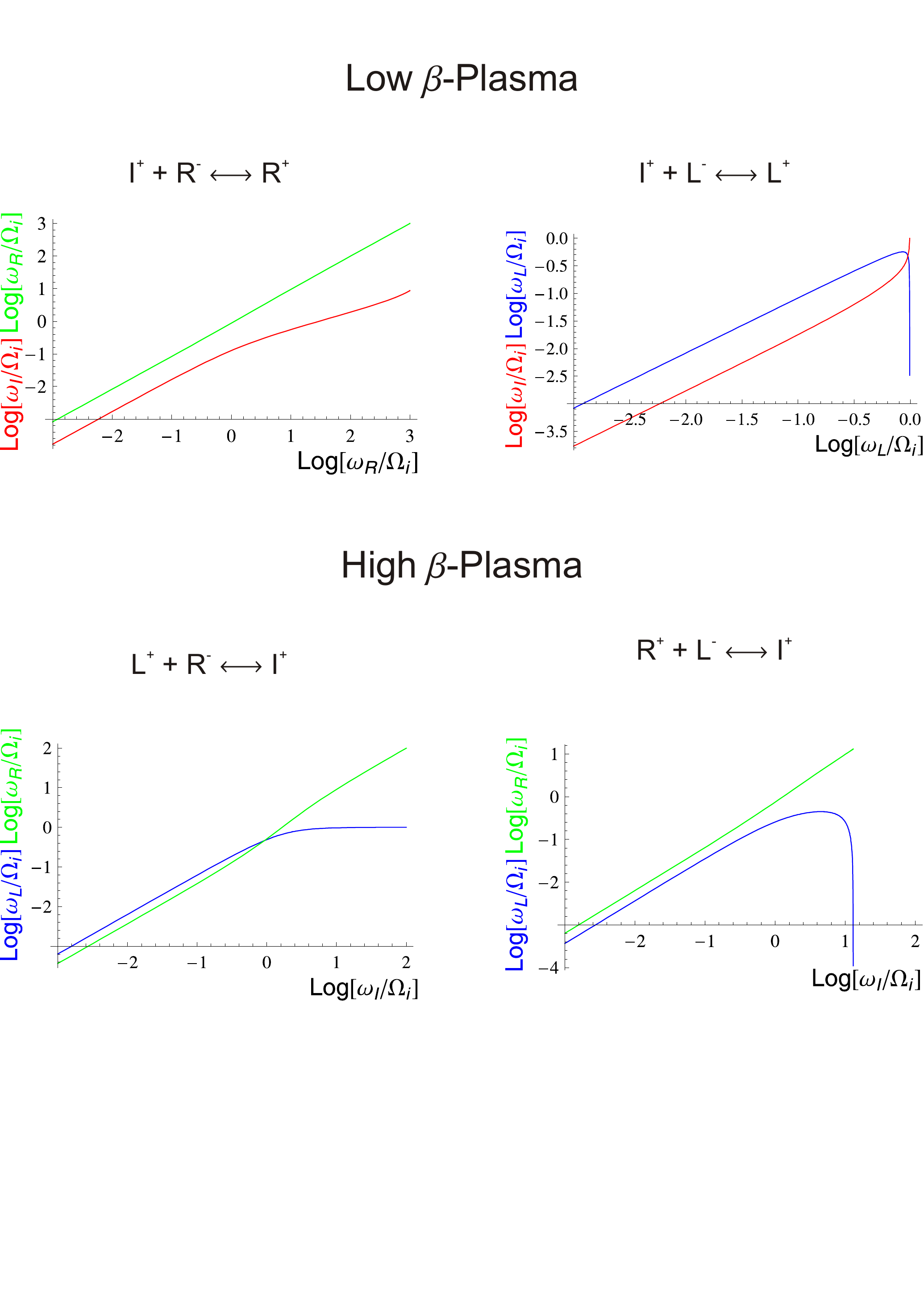}
  \caption{The frequencies of the daughter waves as a function of the
    frequency of the decaying wave in interactions, which fulfill the
    resonance condition in the whole frequency range including the MHD
    regime, analysed earlier in \cite{CW72,VS05}. In the plots L-waves
    (blue [dot-dashed] curve), R-waves (green [dashed] curve) and I-waves (red [solid] curve) can be
    seen.}
  \label{fig:standard}
\end{figure}

In the upper right panel of Fig. \ref{fig:standard} the interaction
$\mathrm{I}^+ + \mathrm{L}^- \leftrightarrow \mathrm{L}^+$ is
shown. This interaction is very similar to the one described in
\cite{VS05}, but for frequencies close to the ion gyrofrequency
$\Omega_\mathrm{i}$ the effect of the L-wave resonance can be seen. At
those frequencies, therefore, an ion-cyclotron wave is able to decay
into a low-frequency Alfv\'en wave and a high-frequency ion sound
wave. This process, however, has to compete against the dissipation of
the mother wave due to ion cyclotron resonance.

The interaction $\mathrm{I}^+ + \mathrm{R}^- \leftrightarrow
\mathrm{R}^+$ (upper left panel of Fig. \ref{fig:standard}) is similar
to the previous one in the low-frequency region, but shows a different
behaviour for $\omega \ge \Omega_\mathrm{i}$, since the R-wave has no
resonance there. The R-waves (the decaying one and the daughter wave)
have a linear frequency relation, while the ion sound wave resembles
the non-linear frequency difference due to the non-linear dispersion
relation of the R-wave. It is noteworthy that the inverse cascading of
the R-waves may proceed all they way from close to the
electron-cycotron frequency through the Whistler range to the MHD
regime.

\begin{figure}
  \pgfimage[width=\textwidth]{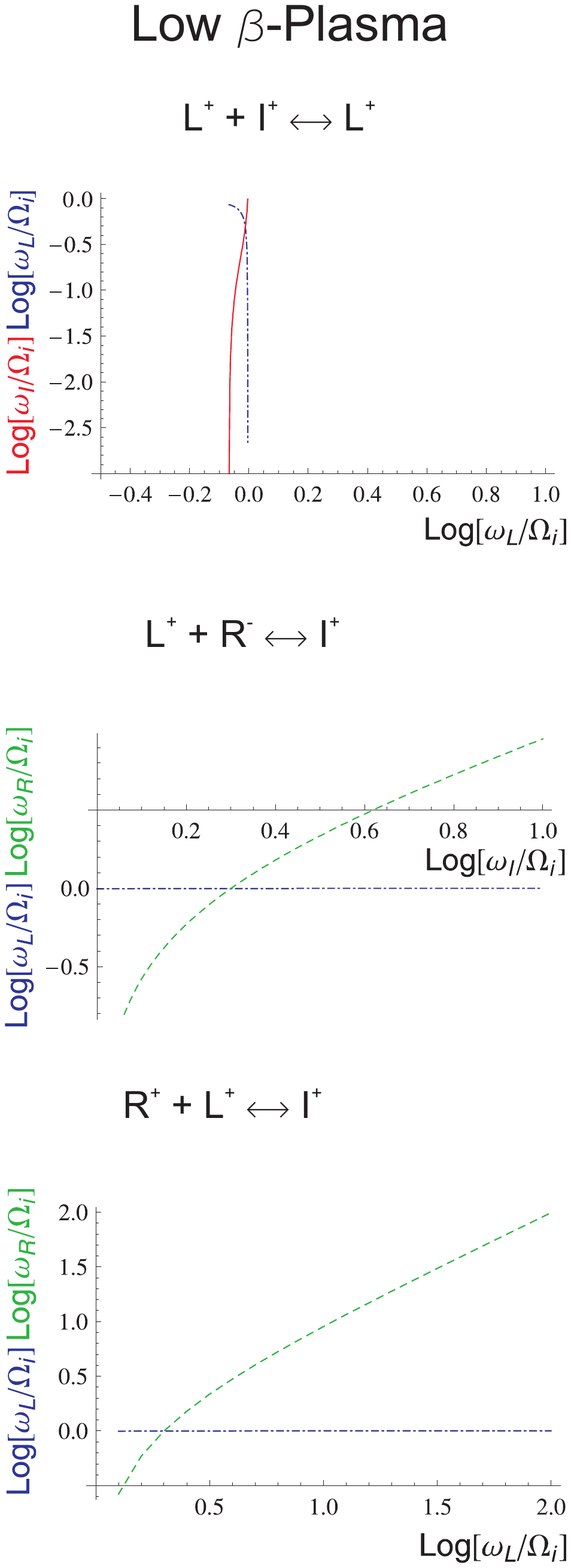}
  \caption{The frequencies of the daughter waves as a function of the
    frequency of the decaying wave in the new interactions, which
    fulfill the resonance conditions only in the dispersive wave
    regime in a low-beta plasma. In the plots L-waves (blue [dot-dashed] curve),
    R-waves (green [dashed] curve) and I-waves (red [solid] curve) can be seen.}
  \label{fig:nonstandard-low}
\end{figure}

In Fig. \ref{fig:nonstandard-low} the limited ranges of validity for
the reactions $\mathrm{L}^+ + \mathrm{I}^+ \leftrightarrow
\mathrm{L}^+$ and $\mathrm{I}^+ \leftrightarrow
\mathrm{L}^++\mathrm{R}^\pm$ are plotted. The dispersive features of
these new interactions are clearly visible.

\subsubsection{High-$\beta$ Plasma}

\begin{figure}
\includegraphics[width=0.8\textwidth]{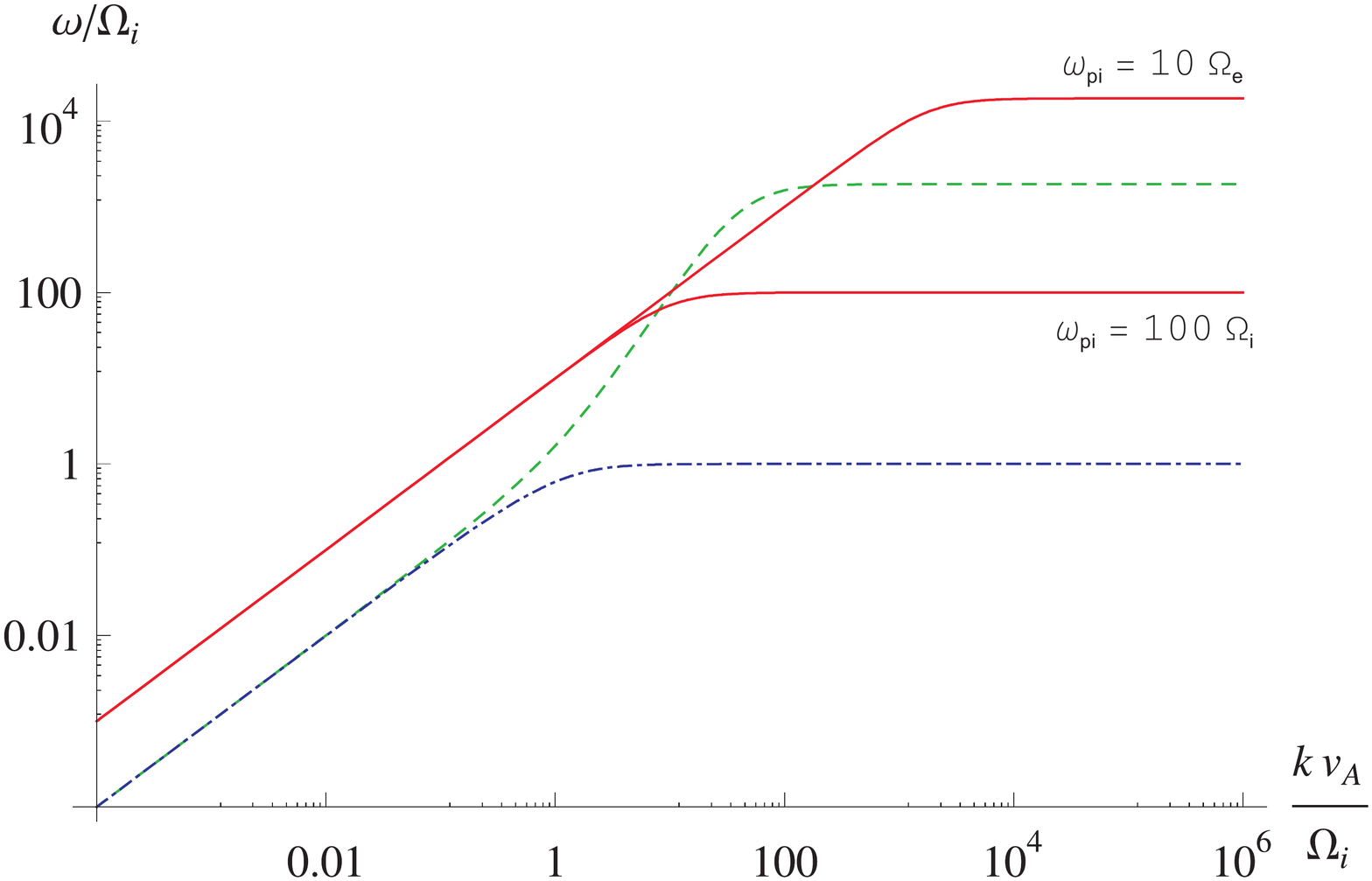}
\caption{Dispersion relation of parallel-propagating plasma waves in
  high-beta plasma, i.e., for $c_\mathrm{s}/v_\mathrm{A}=10$. The
  red (solid) curves represent the ion sound wave for a high-density
  ($\omega_\mathrm{pi}=10\,\Omega_\mathrm{e}$) and low-density
  ($\omega_\mathrm{pi}=100\,\Omega_\mathrm{i}$) plasmas, the blue (dot-dashed)
  curve is the L wave and the green (dashed) curve the R wave.}
\label{fig:disprel2}
\end{figure}

As we have done for the low $\beta$ case, we will also analyse the
interactions with non-vanishing interaction coefficients for a
high-beta plasma.  We will, however, keep the discussion based on
analytics more limited in this case than for the low-beta case. The
dispersion relations in high-beta plasma are depicted in
Fig.~\ref{fig:disprel2}.

For $\beta>1$, the interaction
\begin{eqnarray}
 \mathrm{I}^+ &\leftrightarrow & \mathrm{R}^+ + \mathrm{L}^+
\end{eqnarray}
cannot fulfill the resonance condition at low frequencies, as the
phase speed of longitudinal wave is higher than that of both the
transverse waves. However, if the ion sound speed is below the maximum
value of the whistler phase speed ($\approx 20\,V_\mathrm{A}$) or if
the plasma has a low-enough density so that
$\omega_\mathrm{pi}<\Omega_\mathrm{e}$, the R-wave phase speed exceeds
the sound-wave phase speed (Fig.~\ref{fig:disprel2}) and reactions may
occur.

The interaction
\begin{equation}
\mathrm{L}^+ \leftrightarrow  \mathrm{I}^+ + \mathrm{L}^-
\end{equation}
is not possible in a high-$\beta$ plasma, as
$k_\mathrm{L}(\omega)-\omega/c_\mathrm{s}>0\; \forall\;
\omega<\Omega_\mathrm{i}$ in a high-beta plasma, and this makes it
impossible to fulfill the resonance condition. The interaction
\begin{eqnarray}
\mathrm{L}^+ &\leftrightarrow & \mathrm{I}^+ + \mathrm{L}^+
\end{eqnarray}
will not work either, because $k_\mathrm{L}(\omega) -
\omega/c_\mathrm{s}$ is a monotonically increasing function, which
allows the resonance condition, $k_\mathrm{L}(\omega) -
\omega/c_\mathrm{s}=k_\mathrm{L}(\omega_1) - \omega_1/c_\mathrm{s}$,
to be fulfilled only at $\omega_\mathrm{I}=\omega-\omega_1=0$.

\begin{figure}
  \pgfimage[width=\textwidth]{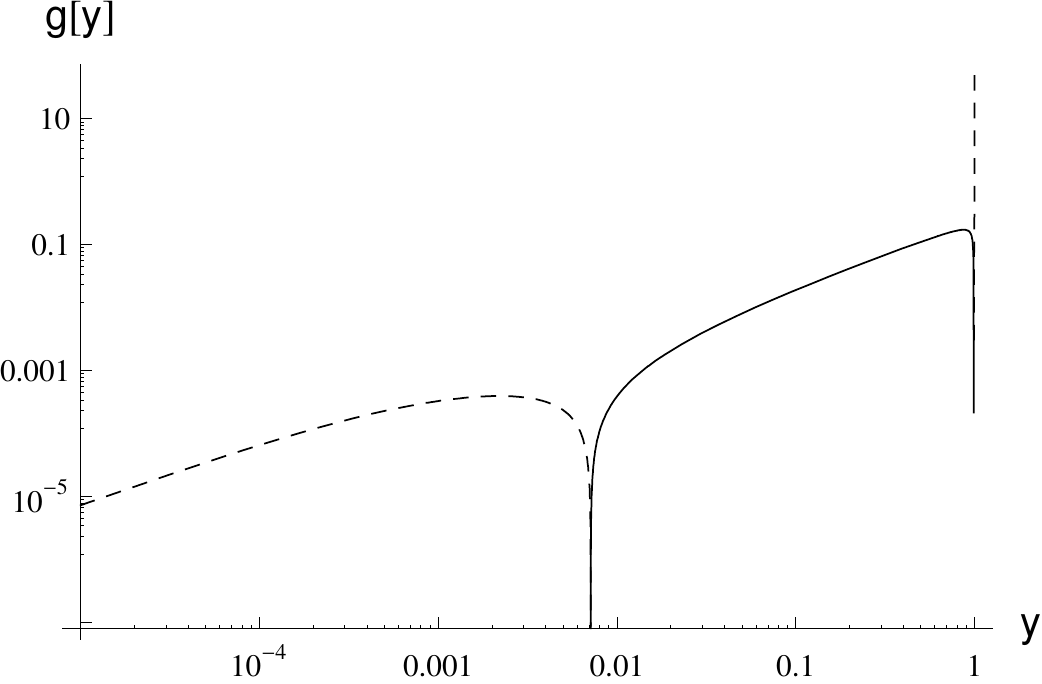}
  \caption{The function of $g(y)$ used to analyse the resonance
    conditions in the interaction
    $\mathrm{R}^+\leftrightarrow\mathrm{R}^++\mathrm{I}^+$. The dashed
    curve gives $|g|$ for $g(y)<0$ and the solid curve the positive
    values of $g$. The value of $\beta$ used in this plot is $\sim
    14$.}
  \label{fig:g}
\end{figure}

However, the interactions
\begin{equation}
\mathrm{R}^+ \leftrightarrow \mathrm{I}^+ + \mathrm{R}^\pm
\end{equation}
are possible as long as at least one of the R waves falls in to the
dispersive frequency range. The resonance conditions for a linear
sound wave, $\omega_\mathrm{I} = c_\mathrm{s}k_\mathrm{I}$, yield
\begin{equation}
k_\mathrm{R}(\omega)-\frac{\omega}{c_\mathrm{s}}=\pm k_\mathrm{R}(\omega_1)-\frac{\omega_1}{c_\mathrm{s}}.\label{eq:RIR}
\end{equation}
The upper sign corresponds to $g(y)=g(y_1)$ with
$y=\omega/\Omega_\mathrm{e}$ and (Fig.~\ref{fig:g})
\begin{equation}
g(y)=\frac{y}{\sqrt{\beta}}-y\sqrt{\frac{m_\mathrm{e}/m_\mathrm{i}}{(m_\mathrm{e}/m_\mathrm{i}+y)(1-y)}},
\end{equation}
which has zeros at $y=0$ and
\begin{eqnarray}
y_\pm&=& \frac{1-m_\mathrm{e}/m_\mathrm{i}\pm\sqrt{(1-m_\mathrm{e}/m_\mathrm{i})^2-4(\beta-1) m_\mathrm{e}/m_\mathrm{i}}}{2}\\
 y_-&\approx&(\beta-1) m_\mathrm{e}/m_\mathrm{i},\quad y_+\approx 1-\beta m_\mathrm{e}/m_\mathrm{i}
\end{eqnarray}
and a single minimum between 0 and $y_-$ and a single maximum between
$y_-$ and $y_+$.  Thus, R waves with frequencies $y$ above the
position $y_\max$ of the maximum are decaying into R waves with
frequencies below $y_\max$, analogously to the low-beta case
$\mathrm{L}^+\leftrightarrow\mathrm{L}^++\mathrm{I}^+$, but now the
mother-wave frequency is very close to $\Omega_\mathrm{e}$. However,
there is also another branch of interactions, which corresponds to the
negative values of $g$. There, R waves with frequencies above the
position $y_\min$ of the minimum of $g$, i.e., with $y\in(y_\min,y_-)$
are decaying into R waves at frequencies $y\in(0,y_\min)$. The value
of $y_\min$ can be found iteratively in a similar manner as $x_\max$
for $f(x)$ above, since at $y<y_-\ll 1$ and with
$x=m_\mathrm{i}y/m_\mathrm{e}$ we have
\begin{eqnarray}
g&=&\frac{m_\mathrm{e}x}{m_\mathrm{i}}\left(\frac{1}{\sqrt\beta}-\frac{1}{\sqrt{1+x}}\right)\\
\frac{\mathrm{d}g}{\mathrm{d}y}&=&\frac{\mathrm{d}x}{\mathrm{d}y}\frac{\mathrm{d}g}{\mathrm{d}x}
= \frac{1}{\sqrt\beta}-\frac{\frac{1}{2}x+1}{(1+x)^{3/2}}
\end{eqnarray}
giving, for $g'(y_\min)=0$, an iterative formula: $x_\min^{(0)}=1$
and
\begin{equation}
x_\min^{(n+1)} = [\beta(\textstyle{\frac{1}{2}}x_\min^{(n)}+1)^2]^{1/3}-1.
\end{equation}
Thus, these solutions are found for mother-wave frequencies
$\omega\in[x_\min\Omega_\mathrm{i},(\beta-1)\Omega_\mathrm{i}]$.

The lower sign in Eq.\ (\ref{eq:RIR}) yields solutions
\begin{equation}
\frac{y}{\sqrt{\beta}}-y\sqrt{\frac{m_\mathrm{e}/m_\mathrm{i}}{(m_\mathrm{e}/m_\mathrm{i}+y)(1-y)}}
=\frac{y_1}{\sqrt{\beta}}+y_1\sqrt{\frac{m_\mathrm{e}/m_\mathrm{i}}{(m_\mathrm{e}/m_\mathrm{i}+y_1)(1-y_1)}},
\end{equation}
meaning that the solutions are given by the roots of
\begin{equation}
g(y)=y_1\left(\frac{1}{\sqrt{\beta}}+\sqrt{\frac{m_\mathrm{e}/m_\mathrm{i}}{(m_\mathrm{e}/m_\mathrm{i}+y_1)(1-y_1)}}\right),
\end{equation}
where the right-hand side is monotonic and can have any value between
$0$ and $\infty$. Thus, this reaction involves mother-wave frequencies
in the whole range where $g(y)$ is positive.

Finally, we are left with two interactions, which have been treated in
\cite{VS05} for the non-dispersive case
\begin{eqnarray}
\mathrm{I}^+ &\leftrightarrow & \mathrm{L}^+ + \mathrm{R}^- \\
\mathrm{I}^+ &\leftrightarrow & \mathrm{R}^+ + \mathrm{L}^-.
\end{eqnarray}
These should work also in the dispersive case. We will analyse these
reactions in more detail numerically.

\paragraph{Numerical analysis.}

As for the low $\beta$-case we plot the solutions for the resonance
condition in the high $\beta$-case. As a model system for a high
$\beta$ plasma we have chosen parameters typical for the downstream
region of a strong coronal shock
\begin{eqnarray}
  \textrm{Number density} && n_i = n_e = 8\times 10^9\, \mathrm{cm}^{-3}\\
  \textrm{Electron temperature} && T_e = 2.5 \times 10^7\, \mathrm{K}\\
  \textrm{Magnetic field} && B = 5 \,\mathrm{G}
\end{eqnarray}
The value of beta for this plasma is
$\beta=c_\mathrm{s}^2/V_\mathrm{A}^2\approx 14$ and
$\omega_\mathrm{pi}\approx 1.3\Omega_\mathrm{e}$.

The first interaction in the high-$\beta$ case has also been discussed
in \cite{VS05}: $\mathrm{I}^+ \leftrightarrow \mathrm{L}^+ +
\mathrm{R}^-$ (lower left panel of Fig. \ref{fig:standard}). The
solution of the resonance condition in the dispersive case is similar
to the non-dispersive case for $\omega \le \Omega_\mathrm{i}$. In the
high-frequency regime we find, that the L-wave goes into
resonance. For low frequencies the interaction takes place for an
L-wave of slightly higher frequency than the R-wave, this changes at
$\omega=\Omega_\mathrm{i}$.  For high frequencies the energy is
supplied almost completely by the R-wave, while the momentum comes
from the L-wave.

The next interaction has also been discussed in \cite{VS05}:
$\mathrm{I}^+ \leftrightarrow \mathrm{R}^+ + \mathrm{L}^-$ (lower
right panel of Fig.~\ref{fig:standard}). Here the dispersive effects
show a completely different behaviour.  The L-wave frequency drops to
zero for high frequencies, while the R-wave shows an almost linear
increase.

\begin{figure}
  \pgfimage[width=\textwidth]{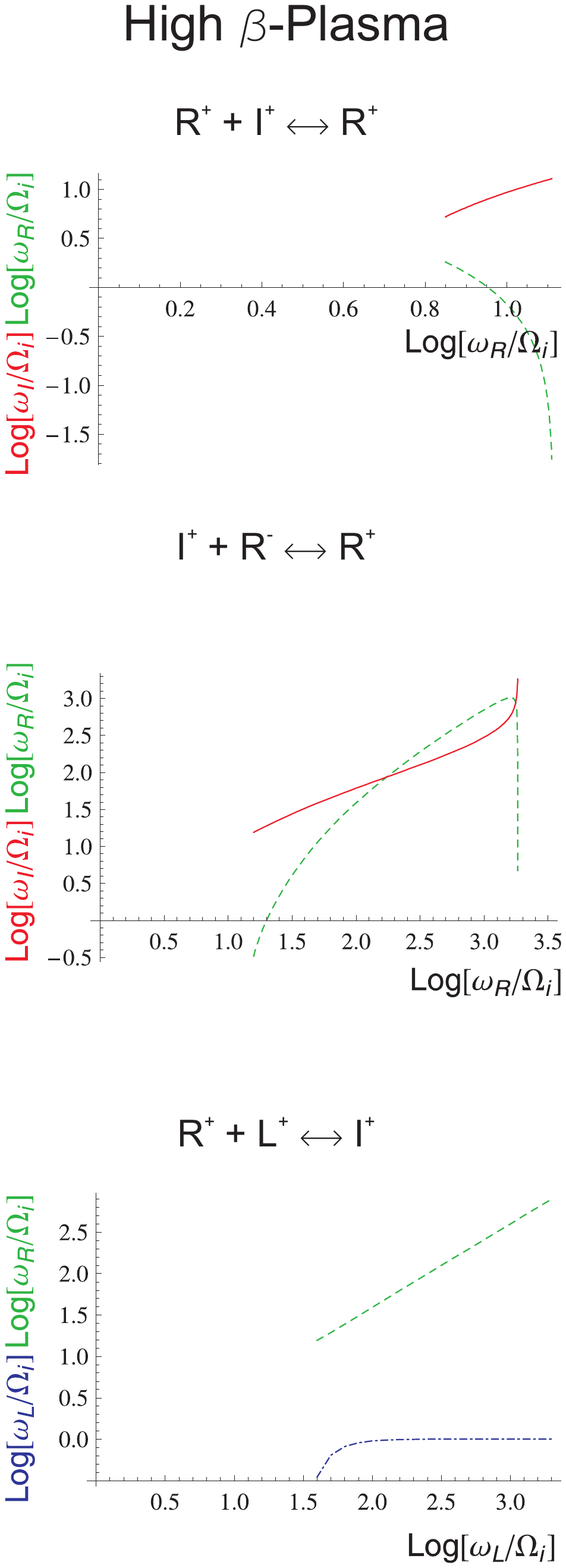}
  \caption{The frequencies of the daughter waves as a function of the
    frequency of the decaying wave in the new interactions, which
    fulfill the resonance conditions only in the dispersive wave
    regime in a high-beta plasma. In the plots L-waves (blue [dot-dashed] curve),
    R-waves (green [dashed] curve) and I-waves (red [solid] curve) can be seen.}
  \label{fig:nonstandard-high}
\end{figure}

For the interaction $\mathrm{R}^+ \leftrightarrow \mathrm{R}^+ +
\mathrm{I}^+$ (top panel of Fig.~\ref{fig:nonstandard-high}) it can be
clearly seen that there is only a small frequency band where the
resonance condition can be fulfilled.  This is due to phase speed of
the R-wave which exceeds that of the sound wave only for a limited
frequency range.

As in the previous case, the interaction $\mathrm{R}^+ \leftrightarrow
\mathrm{I}^+ + \mathrm{R}^- $ (middle panel of Fig. \ref{fig:nonstandard-high})
shows clear dispersive effects. It is limited to the Whistler frequency range
but has a larger range of validity than the previous case. In contrast to the
previous case the upper limit is now given by the electron cyclotron limit of
the R-wave.

Finally, the reaction $\mathrm{I}^ + \leftrightarrow \mathrm{R}^++\mathrm{L}^+$
is plotted in the bottom panel of Fig.~\ref{fig:nonstandard-high}. This shows
that ion-cyclotron waves are able to interact with higher-frequency Whistlers
propagating in the same direction, producing longitudinal waves that are
rapidly damped. This reaction, thus, provides a dissipation mechanism for
Whistlers in plasmas with cold ions and hot electrons.

\section{Discussion and conclusions}
We have analysed three-wave interactions of parallel-propagating plasma waves,
concentrating on the effects of wave dispersion on the interactions. Our
analysis shows that the theory is consistent with the conservation of angular
momentum at microscopic level: the total amount of spin carried by the wave
quanta must be conserved in the three-wave interactions. The reactions
conserving the angular momentum are further analysed by employing the
dispersion relations of left- and right-hand circularly polarised waves at
frequencies below $\Omega_\mathrm{i}$ and $\Omega_\mathrm{e}$, respectively,
and of ion-sound waves at frequencies below $\omega_\mathrm{pi}$.

Our analysis extends the previously analysed \cite{CW72,VS05} low-beta
interactions,
\begin{eqnarray*}
\mathrm{L}^+&\leftrightarrow&\mathrm{I}^++\mathrm{L}^-\\
\mathrm{R}^+&\leftrightarrow&\mathrm{I}^++\mathrm{R}^-
\end{eqnarray*}
to higher frequencies and shows that the latter reaction may occur also in
high-beta plasma at Whistler frequencies of the mother R wave, where its phase
speed exceeds that of the ion sound wave. The implied efficient inverse
cascading of the R waves has important consequences on energetic particle
transport and acceleration in collisionless plasmas: as higher-energy particles
generally interact with lower-$k$ waves, this mechanism allows waves generated
by low-energy particles to be converted to waves resonant with high energies,
and this will enhance the rate of stochastic acceleration of the electrons, for
example.  Also the theory of diffusive shock acceleration (DSA) greatly
benefits from an inverse cascade: it is the wave intensities at the lowest
frequencies that determine the maximum energies produced in a shock in a given
time, but as the upstream spectrum of plasma waves is usually generated by
streaming instabilities due to the accelerated particles themselves, the
spectrum of waves is usually an increasing function of frequency in the range
resonant with the highest energy particles \cite{VL07}. Obviously, an inverse
cascade will increase the intensity of waves resonant with the highest energy
particles and, therefore, increase the rate of acceleration.

The previously analysed high-beta interactions,
\begin{eqnarray*}
\mathrm{I}^+&\leftrightarrow&\mathrm{R}^++\mathrm{L}^-\\
\mathrm{I}^+&\leftrightarrow&\mathrm{L}^++\mathrm{R}^-
\end{eqnarray*}
are also extended to higher frequencies. Both of these reactions are also
possible in low-beta plasmas, as long as the frequency of the L wave is close
to the ion-cyclotron frequency. As sound waves are rapidly damped in general,
the reactions usually proceed to the direction of wave coalescence. Thus, these
reactions may represent a new dissipation channel to the R waves in low-beta
plasmas, although cyclotron damping of the L waves is also large near the
resonance.

We have also discovered completely new interactions, which are not possible in
an MHD description. In these reactions, all three waves propagate in the same
direction,
\begin{eqnarray*}
\mathrm{I}^+&\leftrightarrow&\mathrm{R}^++\mathrm{L}^+\\
\mathrm{L}^+&\leftrightarrow&\mathrm{L}^++\mathrm{I}^+\quad(\mbox{low }\beta)\\
\mathrm{R}^+&\leftrightarrow&\mathrm{R}^++\mathrm{I}^+\quad(\mbox{high }\beta),
\end{eqnarray*}
which is strictly forbidden for non-dispersive waves. The latter two
interactions shift energy from the dispersive frequency range to the MHD range,
but an inverse cascade does not really develop, as the wave falls out of the
frequency range able to decay already after the first interaction. This
process, nevertheless, may help us understand, why spectrum of magnetic
fluctuation in the solar wind experiences a clear break above the ion-cyclotron
frequency \cite{D83}, although the right-handed mode should not experience
cyclotron damping in this frequency range.

Assuming that the sound waves are rapidly damped, all the discussed
three-wave interactions provide a means of plasma heating. As the
dissipation mechanism is not based on cyclotron resonance, the
resulting heating rates on ions and electrons may be significantly
different from direct ion-cyclotron damping.

In this paper, we only made use of the symmetries of the response
tensor to find out the interactions with non-zero interaction
rates. In a follow-up paper, we will study numerically the interaction
rates of the three-wave interactions presented in this paper. This
will allow us to deduce the importance of the interactions relative to
other plasma phenomena, and determine the effects of three-wave
interactions on plasma heating rates and particle acceleration.

In future work, it will also be important to extend the study also to waves
propagating at a finite angle with respect to the magnetic field. In this case,
as the polarisation tensor will become more complicated and the response tensor
loses its transverse isotropy, many reactions should be allowed to occur.
Another potentially important point to include to the model is multiple ion
species. As the L waves will develop more resonances and subsequent cutoffs,
especially the interactions $\mbox{L}^+\to \mbox{I}^++\mbox{L}^\pm$ will be
different from the ones presented here.

In conclusion, three-wave interactions in collisionless plasmas
involving dispersive waves show a richer variety of possible reactions
than the MHD counterparts. These reactions are potentially important
in many branches of plasma astrophysics ranging from particle
acceleration to plasma heating. We have identified some of them, but
the theory has many more application not discussed in this paper.

\section*{Acknowledgements}

FS acknowledges support from the Deutsche Forschungsgemeinschaft through grant
SP 1124-1/1.


\newpage


\begin{thebibliography}{99}
\bibitem{M86} D. Melrose, Instabilities in Space and Laboratory
  Plasmas, Cambridge University Press, Cambridge \textbf{(1986)}

\bibitem{CW72} Y.-C. Chin and D. G. Wentzel, Astrophysics and Space
  Sciences 16, 465--477 \textbf{(1972)}

\bibitem{W74} D.G. Wentzel, Solar Physiscs 39, 129--140 \textbf{(1974)}

\bibitem{S75} J. Skilling, Monthly Notes of the Royal astronomical Society 173, 255--269 \textbf{(1975)}

\bibitem{VS05} R. Vainio and F. Spanier, Astronomy and Astrophysics
  437, 1--8 \textbf{(2005)}

\bibitem{LM06} Q. Luo and D. Melrose, Monthly Notes of the Royal
  astronomical Society 368, 1151--1158 \textbf{(2006)}

\bibitem{MS72a} D. Melrose and W. Sy, Astrophysics and Space Sciences 17,
  343--356 \textbf{(1972)}

\bibitem{MS72b} D. Melrose and W. Sy, Australian J. Physics, 25,
  387--402 \textbf{(1972)}

\bibitem{S62} T.H. Stix, The Theory of Plasma Waves, McGraw-Hill, New
  York\textbf{(1962)}

\bibitem{VL07} R. Vainio and T. Laitinen, Astrophysical J., 658,
  622--630 \textbf{(2007)}

\bibitem{D83} K.U. Denskat, H.J. Beinroth, and F. M. Neubauer, 1983,
  J. Geophysics, 54, 60--67\textbf{(1983)}

\end{thebibliography}
\end{document}